\documentclass[%
 aip,
 amsmath,amssymb,
 reprint,%
]{revtex4-1}

\usepackage{graphicx}
\usepackage{dcolumn}
\usepackage{bm}
\usepackage{amsmath, amssymb, stmaryrd, xparse, mleftright}
\usepackage{color}
\usepackage{array}
\usepackage{ulem}
\usepackage[utf8]{inputenc}
\usepackage[T1]{fontenc}
\usepackage{mathptmx}
\usepackage{etoolbox}

\usepackage{blindtext, xcolor}
\usepackage{xcolor}
\usepackage{hyperref}
\hypersetup{
    colorlinks,%
    citecolor=blue,%
    linkcolor=blue,%
    urlcolor=blue
}

\begin{document}

\preprint{AIP/123-QED}

\title{Weyl semimetal engineering by symmetry control in NiTe$_2$} 

\author{Marcos G. O. Junior} 
\affiliation{Instituto de Física, Universidade Federal de Uberlândia, Uberlândia, MG 38400-902, Brazil}

\author{Augusto L. Araújo}
\altaffiliation{Present address: Ilum School of Science, Brazilian Center for Research in Energy and Materials (CNPEM), 13083-970, Campinas, SP, Brazil}
\affiliation{Instituto de Física, Universidade Federal de Uberlândia, Uberlândia, MG 38400-902, Brazil}

\author{Emmanuel V. C. Lopes}
\thanks{Author to whom correspondence should be addressed: \href{mailto:emmanuel.lopes@ufu.br}{emmanuel.lopes@ufu.br}}\affiliation{Instituto de Física, Universidade Federal de Uberlândia, Uberlândia, MG 38400-902, Brazil}

\author{Tome M. Schmidt}
\thanks{Author to whom correspondence should be addressed: \href{mailto:tschmidt@ufu.br}{tschmidt@ufu.br}}\affiliation{Instituto de Física, Universidade Federal de Uberlândia, Uberlândia, MG 38400-902, Brazil}

\begin{abstract}

In this work, we investigate the emergence of Weyl points in an inversion symmetry-breaking 1T-NiTe$_2$ system. Through first-principles calculations based on the density functional theory combined with tight-binding methods, we find three distinct sets of Weyl crossings under an appropriate symmetry breaking. The first set, composed of four Weyl points, emerges from the Dirac semimetal. Surprisingly, the other two sets result in additional twenty-four Weyl crossings, depending on the weight of the symmetry breaking. We investigate the topological characteristics of the Weyl semimetals by computing the Weyl chirality, Berry curvature, and the evolution of Wannier charge centers. Additionally, the bulk-boundary correspondence has been shown by computing the Fermi arcs. Our results provide a way for creating and manipulating distinct sets of Weyl points with appropriate external control, which can be valuable for applications in Weyltronics.

\end{abstract}

\maketitle

The Weyl fermions were theoretically predicted by Hermann Weyl in 1929~\cite{Weyl-0} as one of the particular cases of the Dirac equation. These particles behave as massless relativistic fermions, which arise in pairs with opposite chiralities.\cite{Armitage2018, Felser2017, Hasan2017} Although predicted in high-energy physics, the only experimental observation of Weyl fermions remains in solid state physics, where they arise in the materials context under either time-reversal or inversion symmetry breaking.\cite{SYang2016, Hasan2017, Qi2020} Usually, Weyl crossings come from symmetry-protected Dirac semimetals (DSMs), with crossing points that are fourfold degenerate. Symmetry breaking raises the fourfold degeneracy, resulting in a pair of Weyl crossings with a defined chirality. \cite{Murakami2007, Burkov2011, Burkov2018, Schmidt2020} Due to the Nielsen-Ninomiya theorem, the sum of chiralities over the Brillouin zone (BZ) must vanish.\cite{Nielsen1983} Although Weyl points (WPs) usually emerge from DMs, they can also arise from gapped regions.\cite{Lu2024, LiuTCI2019, Wang2019, Chang2016}
 
Similarly to other topological phases of matter, driven by the bulk-boundary correspondence, Weyl semimetals also exhibit topological surface states, which are called Fermi arcs. Unlike other topological classes, where the surface states connect the valence to conduction bands, the Fermi arcs perform a non-closed curve in momentum space, connecting the bulk WPs with opposite chiralities projected onto the surface.\cite{Hasan2017, Felser2017} Noncentrosymmetric XY compounds (X = Ta, Nb and Y = As, P) represent the first class of Weyl semimetal materials with Fermi arc properties that were theoretically predicted and experimentally observed.\cite{Hasan2015, Xu2015Niob,XuTaP2015, Yang2015, Lv2015, Liu2015} Later, Weyl crossings have also been experimentally reported in additional noncentrosymmetric systems, such as MoTe$_2$~\cite{Sun2015, Jiang2017, Wang2016prl, tamai2016} and WTe$_2$,~\cite{Wu2016, Wang2016} as well as in magnetic materials.~\cite{Liu2019, Morali2019, Schrunk2022, Schrter2020} Weyl semimetals hold interesting properties, such as protected surface states,\cite{Hasan2017, Armitage2018} negative magnetoresistance,\cite{HuangPRX2015} and chiral anomaly.\cite{Ong2021, Zhang2016} These properties not only make this class of topological materials promising for applications in different subfields of condensed matter physics, including catalysis,\cite{Sudrajat2025}  thermoelectricity,\cite{Han2020, Kozii2019} and superconductivity,\cite{Weyl-19,Weyl-20,Weyl-21, Sukhachov2020} but it is also a platform to realize and understand the concept of fundamental particles.

A particular class of materials promising for the study of Dirac semimetals is the transition metal dichalcogenides (TMDs). The TMD system has the form MX$_2$, where M represents a transition metal, which is encapsulated by chalcogen atoms, denoted by X.\cite{Manzeli2017, Han2018} These systems can crystallize in the 1T, 1T', and 2H structures, and the deposited layers are held together by van der Waals (vdW) interactions. Many examples of TMDs have been reported to present DSM phase and, in some cases also Weyl properties, such as MTe$_2$ (M = Mo, W, Pd and Pt) \cite{Sun2015,Leng2017, Yan2017,Wu2016, Wang2016} and PtSe$_2$.\cite{Fischer2024, Huang2016} Another interesting TMD is the centrosymmetric 1T-NiTe$_2$, which has been widely studied and, through \textit{ab inito} calculations and angle-resolved photoemission spectroscopy (ARPES) measurements, it has been verified to belong to the Dirac semimetal phase.\cite{Gosh2019, Zhang2020, Zhang2021, Chiang2021}
 
In this work, using first-principles and topological invariant calculations, we investigate the formation of multiple Weyl points in NiTe$_2$ under inversion symmetry breaking. We find that, by breaking the mirror $\sigma_{xy}$ symmetry, the fourfold degenerate Dirac crossing splits into a pair of Weyl points with opposite chiralities near the Fermi level. Notably, beyond the two pairs of WPs originated from the Dirac semimetal phase, we find two additional sets of WPs coming out from gapped regions. The emergence of these additional sets depends not just on the symmetry, but also on the weight of the symmetry-breaking, resulting in additional 6 or 12 new pairs of WPs throughout the BZ. The topological nature of these additional WPs was verified by nonvanishing Berry curvature, Weyl chirality calculations, as well as their corresponding Fermi arcs. Our results provide a way to create sets of Weyl points in NiTe$_2$ by symmetry control.

First-principles calculations were performed using the density functional theory (DFT) \cite{dft1, dft2} within the generalized gradient approximation (GGA) for the exchange and correlation functional, employing the Perdew-Burke-Ernzerhof (PBE) parametrization \cite{PBE}. A fully relativistic pseudopotential, within the projector augmented wave method (PAW),\cite{PRBblochl1994, Corso2010} was used in the noncollinear spin-DFT formalism self-consistently. We have used the DFT codes Vienna \textit{ab initio} Simulation Package (VASP) \cite{vasp1, vasp2} and Quantum Espresso (QE),\cite{Giannozzi2009,Giannozzi2017} with plane wave basis set with a cut-off energy of 400 eV. The Brillouin zone was sampled with a 25{$\times$}25 {$\times$}25 $k$-mesh grid (Monkhorst-Pack scheme) such that the total energy converges within the meV scale, and atomic structures are optimized requiring that the force on each atom to be less than $0.01$ eV/{\AA}. The analysis of the DFT results, including band structure plots, projections of orbital-atomic contributions, and isosurfaces plots, has been performed using the VASProcar post-processing code.\cite{vasprocar} We build a tight-binding model based on DFT results by means of the Wannier90 package.\cite{Wannier90} The topological properties, such as Chern number, Berry curvature, and Fermi arcs were investigated with Wanniertools.\cite{WannierTools}

\begin{figure}[htb]
    \centering
    \includegraphics[width=\columnwidth]{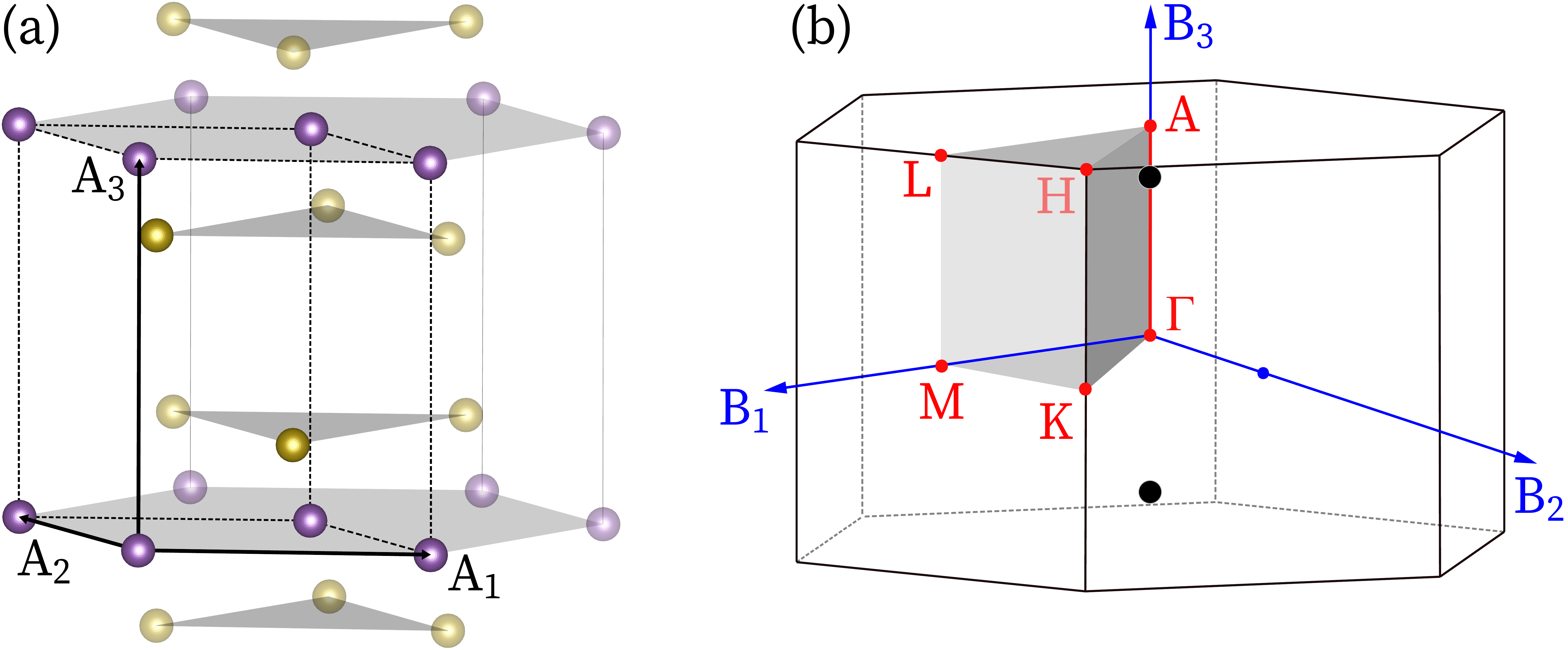}
    \caption{(a) 1T-NiTe$_2$ crystal structure, identifying the unit cell with dashed lines. The purple and golden spheres represent the Ni and Te ions, respectively. The first Brillouin zone is shown in (b), where the black dots denote the DSM locations.}
    \label{fig:Lattice_BZ}
\end{figure}

1T-NiTe$_2$ is a semimetallic TMD, with strong intralayer bonds and vdW interactions between the layers. This system shows a hexagonal crystal structure, as illustrated in Fig.~\ref{fig:Lattice_BZ}(a), with our fully optimized lattice parameters $a=b=3.90$ {\AA} and $c=5.24$ {\AA}. It belongs to the D$^{3}_{3d}$  (or P$\overline{3}$m1) space group symmetry, with time-reversal and inversion symmetry preserved. In this way, due to the Kramers degeneracy, the band dispersion must be doubly degenerate, and any crossing point would be at least fourfold degenerate.  If such a crossing has linear dispersion, it is a strong candidate to be a DSM.\citep{li2020topological} In the NiTe$_2$ system, a fourfold degenerate Dirac crossing has been reported along the $\Gamma$-A path,\cite{Zhang2020, Chiang2021,Fischer2024} as indicated in Fig.~\ref{fig:Lattice_BZ}(b). The high symmetry points $\Gamma$ and A show the full D$_{3d}$ symmetry, while the line connecting the points has a lower symmetry, C$_{3v}$.

\begin{figure}[th]
    \centering
    \includegraphics[width=\columnwidth, height=13.0 cm]{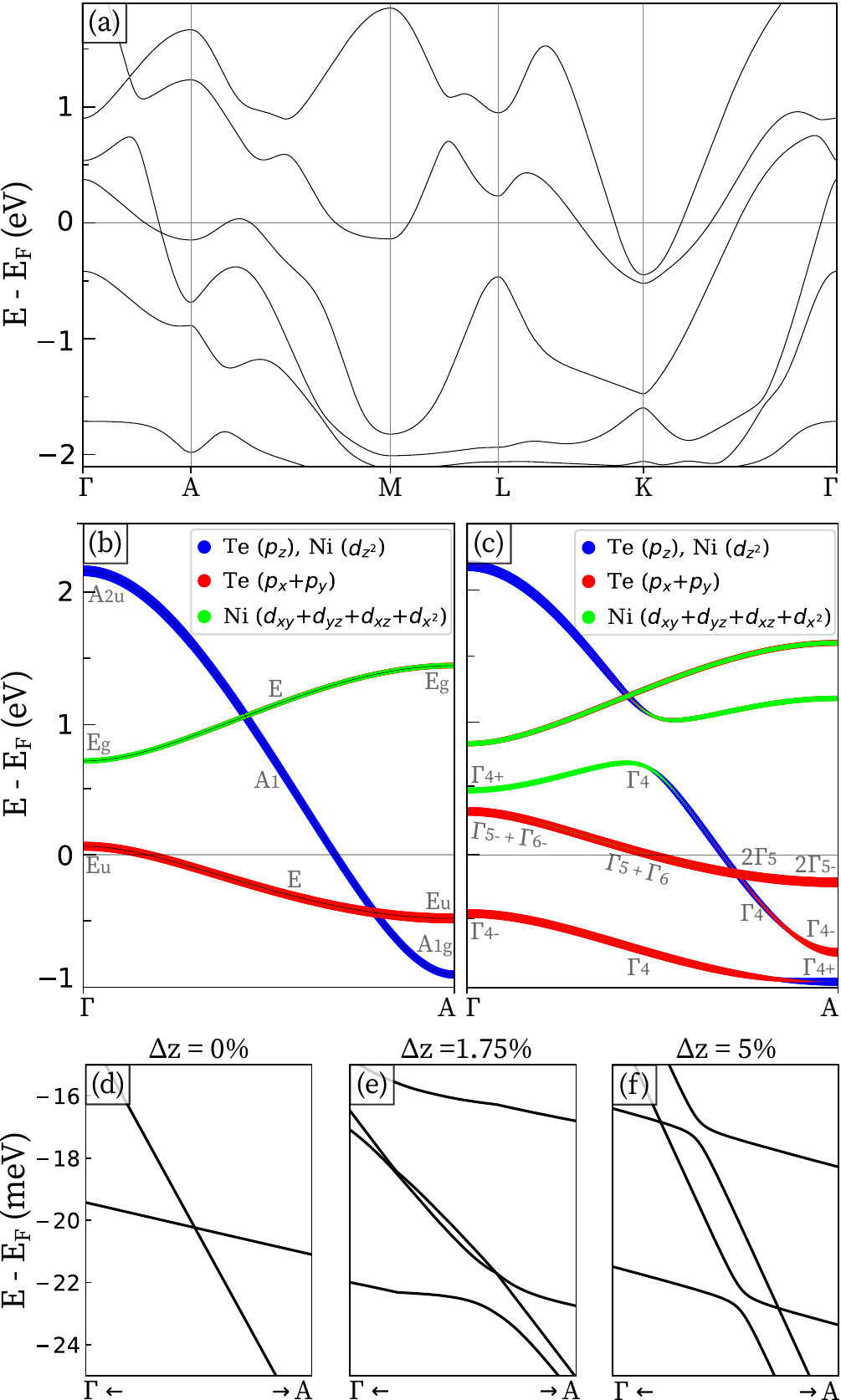}
    \caption{1T-NiTe$_2$ electronic band structure (a). Orbital projection from bands around the Fermi level along the $\Gamma$-A direction without (b) and with SOC (c). The bands are labeled with their corresponding IRREPs. In (d) is highlighted the Dirac semimetal in NiTe$_2$, in (e) and (f) is shown its splitting into Weyl points under distinct inversion symmetry breaking weight.}
    \label{fig:Bands_and_Projections}
\end{figure}

The band structure of NiTe$_2$, shown in Fig.~\ref{fig:Bands_and_Projections}(a), presents a linear crossing point close to the Fermi energy (-0.02 eV) along the $\Gamma$-A direction, a typical type-II DSM. It is interesting to note that, without SOC [Fig.~\ref{fig:Bands_and_Projections}(b)], this band crossing is sixfold degenerate. The inclusion of spin-orbit coupling (SOC) lifts the band degeneracy at the crossing point, resulting in a fourfold degeneracy, as can be seen in  Fig.~\ref{fig:Bands_and_Projections}(c). The DSM phase is intrinsic as a consequence of the symmetry protection, since there is no phase transition. Due to the SOC interaction, a double group symmetry analysis shows not only changes in the atomic orbital character, but also an inversion of the  $\Gamma_{4+}$ and $\Gamma_{4-}$ bands at the high-symmetry A point. This inversion leads to the Dirac semimetal topological phase. The two-dimensional irreducible representations (IRREPs) of the states at the Dirac crossing $\Gamma_{n-}$ and $\Sigma_{m-}$ ($m$, $n$ integers) are characterized by an opposite sign under C$_3$ rotation.\citep{koster} In this way, the C$_3$ symmetry avoids interaction between these states, preventing band gap opening and consequently imposing a Dirac crossing. As the system is semimetal, we also confirmed the DSM topological phase by topological quantum chemistry analysis, in which the real space symmetry of the bands is included besides the topological properties in momentum space.\citep{PhysRevX.7.041069, TQC1,TQC2}

\begin{figure*}[th]
    \centering
    \includegraphics[width=2.08\columnwidth]{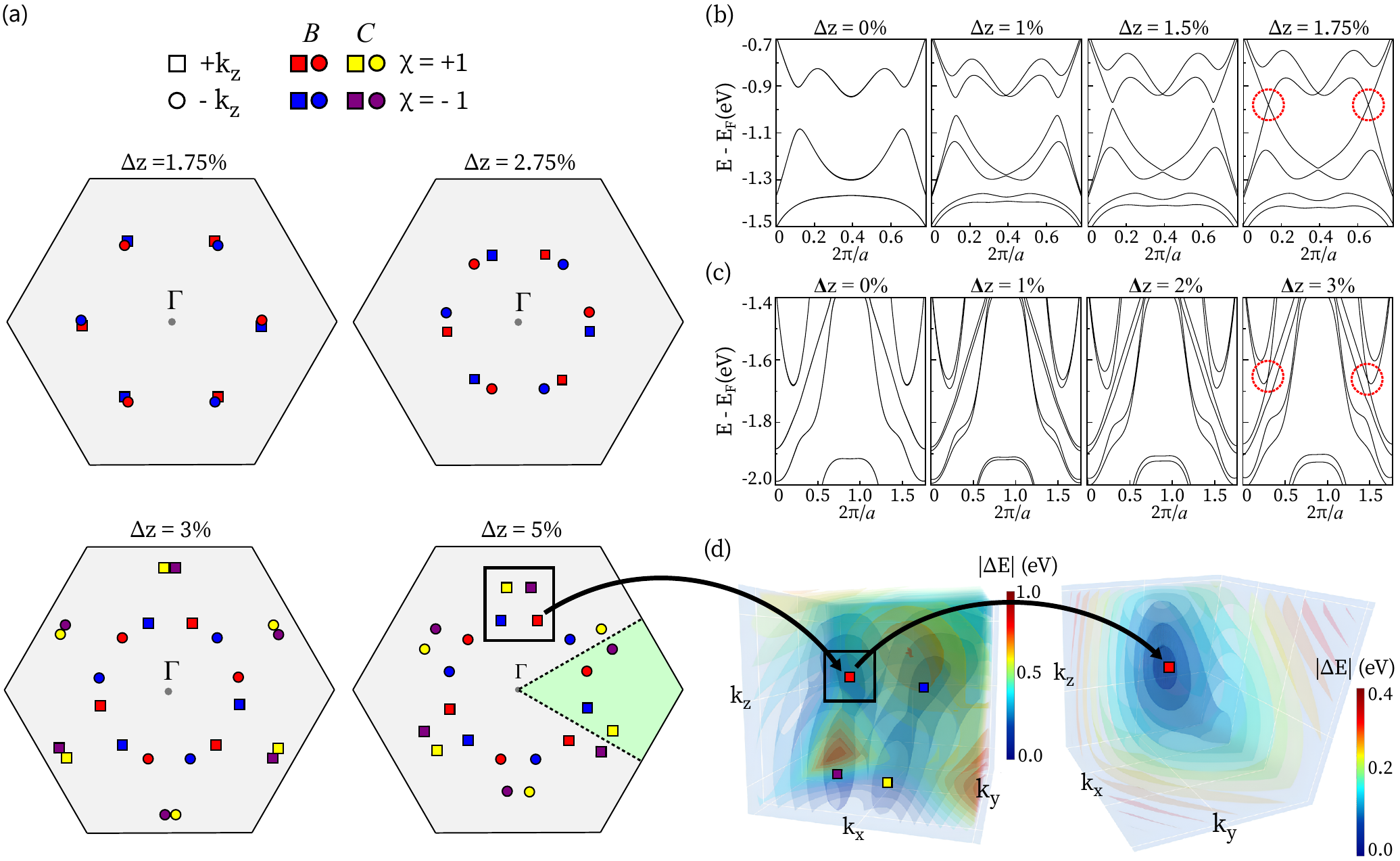}
    \caption{(a) Evolution of Weyl point locations as a function of the symmetry breaking weight ($\Delta{z}$) for sets $B$ (blue and red) and $C$ (yellow and purple). The band structures in (b) and (c) show the emergence of Weyl crossings of sets $B$ and $C$, respectively. The corresponding isoenergy surface plot, illustrating the Weyl point positions (d).}
    \label{fig:distorcao_WPs}
\end{figure*}

The degeneracy of the DSM can be lifted upon breaking either inversion or time-reversal symmetry, allowing the emergence of Weyl semimetals. However, 1T-NiTe$_2$ is a centrosymmetric crystal, unlike other TMDs that are intrinsically noncentrosymmetric, such as 1T'-MoTe$_2$ and T$_d$-WTe$_2$.\cite{Jiang2017, Wang2016prl, tamai2016,Wu2016, Wang2016} To achieve a Weyl semimetal phase, we break the inversion symmetry, keeping the C$_3$ rotation, since it protects gapping, as previously discussed. In a practical way, this can be done by applying an electric field along the $z$ direction of the structure in Fig.~\ref{fig:Lattice_BZ}(a), or by the effect of a substrate, which also breaks the mirror $\sigma_{xy}$ symmetry. The electric field is simulated here by dislocating the Ni atomic plane in relative percentages of the $c$ lattice parameter ($\Delta z$). This correspondence has been verified by computing the average potential energy in a single layer, as shown in Fig. S1. The $\Delta{z}$ atomic displacement induces a potential difference ($\Delta${V$_{z}$}) between the Te atoms [Fig. S1(b)], leading to a built-in electric field. Our results show that the potential energy difference ranges from 0.32 eV to 1.65 eV for Ni displacements in the range between 1\% and 5\%, corresponding to electric fields of $0.12$ eV/{\AA} and $0.65$ eV/{\AA}, respectively. This equivalence of electric field with Ni displacement is illustrated by applying an external electric field of  $0.65$ eV/{\AA} in the system. As shown in Fig. S1(c), the potential energy difference between the tellurium atoms matches the result of Ni atomic displacement.

The symmetry of the point group will be reduced from D$_{3d}$ to C$_{3v}$. As shown in Figs.~\ref{fig:Bands_and_Projections}(d)-(f), the fourfold DSM breaks into a pair of Weyl points aligned along the $k_z$ direction. It preserves the type-II character, the same of the DSM. From Figs.~\ref{fig:Bands_and_Projections}(e)-(f), we note that the Weyl pairs distance increases by increasing the symmetry-breaking weight. The energy splitting between the Weyl pairs is only a few meV, and their chirality WPs distance is less than 0.005\% of the reciprocal lattice parameter, making them hard to be detected. The respective Weyl points are located at (0.00000631, -0.00002383, 0.36463530) and (-0.00000971, 0.00001055, 0.36456357) in units of (2$\pi/a$, 2$\pi/a$, 2$\pi/c$), for $\Delta{z}=5\%$. The other two Weyl points from this set, henceforth denoted as set \textit{A}, can be obtained by applying the time-reversal symmetry operation, which in a 3D system keeps the same chirality.\cite{Vanderbilt2018} 

Weyl points from Dirac semimetals can also be obtained by breaking time-reversal symmetry. Recently, it has been shown that Cr-doped NiTe$_2$ induces ferromagnetic order, which leads to spin splitting in the Dirac semimetal, originating Weyl points near the Fermi energy.\cite{Zhang2024}

By breaking the inversion symmetry, we also verify the emergence of two additional sets of Weyl nodes. These crossing points arise from gapped regions under different weight of symmetry breaking, as illustrated in Fig. \ref{fig:distorcao_WPs}(a). We denote the additional sets of Weyl points as sets \textit{B} and \textit{C}. In the set \textit{B}, represented by blue and red symbols, the topological phase transition from a  trivial gapped region to a Weyl semimetal occurs at $\Delta{z} = 1.75\%$ [Fig. \ref{fig:distorcao_WPs}(b)]. It can be seen pairs of linear touching points occurring at symmetric positions, typically a type-I Weyl semimetal.

By increasing the crystal field distortion, the set \textit{C} emerges at $\Delta{z}$ $= 3\%$, as can be seen in Fig.~\ref{fig:distorcao_WPs}(c). In this picture, we observe a tilted crossing behavior that is a key signature of type-II Weyl crossings. It is worth noting that both additional sets of WPs (sets $B$ and $C$) emerge from the same bands of the set $A$, the one coming from the Dirac semimetal. For the highest distortion investigated in this work ($\Delta{z} = 5\%$), the crossing points of sets \textit{B} and \textit{C} are localized below the Fermi level at -0.89 eV and -1.41 eV, respectively.

The weight for the displacement $\Delta{z}$ affects the WP energy as well as its position in the BZ.
For the highest distortion [Fig~\ref{fig:distorcao_WPs}(a)], we have four WPs in the first BZ (denoted by the green marked region), two of them come from the set $B$ and two from set $C$. The coordinates of the set $B$, in units of (2$\pi/a$, 2$\pi/a$, 2$\pi/c$), are (0.2348, -0.1844, 0.3283) and (0.2332, -0.04809, -0.3321), indicated by the blue square and red circle, respectively. For set $C$, in the same units, the respective coordinates are (0.3245, -0.0305, -0.2212) and (0.3239, -0.2928, 0.2217), indicated by the purple circle and yellow square. The other WP coordinates can be obtained from these four WPs by applying the following symmetries: C$_3$, time-reversal and vertical mirror. While C$_3$ and time-reversal leads to a WP with the same chirality, the vertical mirror reverse the chirality of each WP.

The WPs have also been investigated by scanning any crossing bands in the first BZ. This analysis ensures that we count all Weyl nodes. The Weyl point positions can be determined by computing the energy isosurface $|\Delta E| =|E_{c} - E_{v}|$ between the bands around the Fermi level, within the entire BZ. Here $E_C$ and $E_V$ denote the two bands that originate the Dirac semimetal, as shown in Fig.~\ref{fig:Bands_and_Projections}. For a touching point, this difference will go to zero. In Fig.~\ref{fig:distorcao_WPs}(d) we show a projection of $|\Delta E|$ around a square volume containing a pair of WPs from each distinct set. As can be seen in the zoomed picture, $|\Delta E|$ goes to zero at the crossing point.

\begin{figure}
    \includegraphics[width=8.5 cm, height=6 cm]{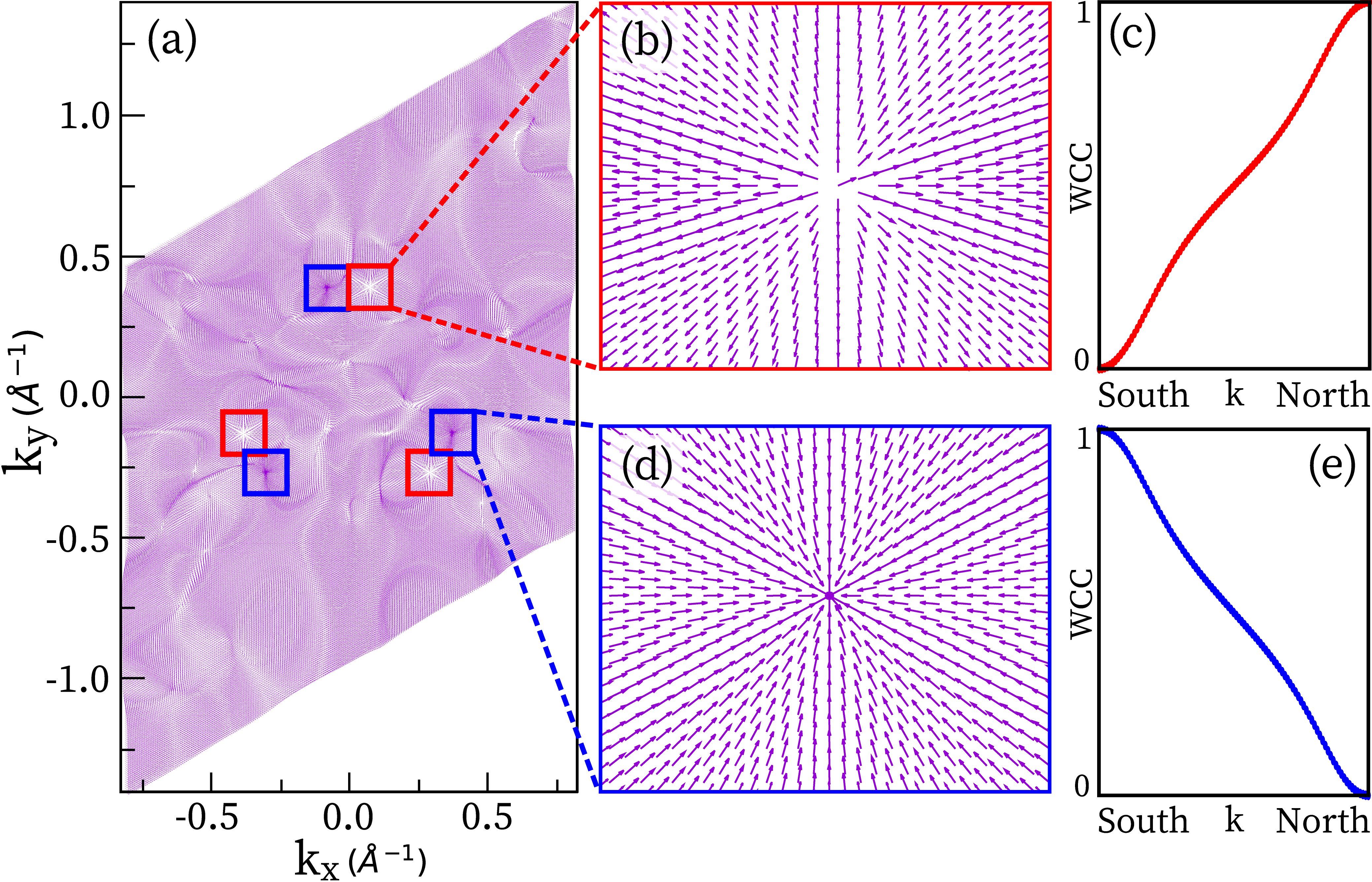}
    \caption{Berry curvature plot for a constant $k_z$ plane in the first Brillouin zone (a). Zoom of two Weyl points illustrating the source (b) and drain (d) of Berry curvature, and their corresponding evolution of WCCs in (c) and (e), respectively. }
    \label{fig:WPS_berry}
\end{figure}

To investigate the topological properties of each Weyl crossing candidate, we calculate the Weyl chirality ($\chi$). For each crossing point, our results show $\chi = \pm1$. The chiralities are shown in Figure ~\ref{fig:distorcao_WPs}(a), illustrated by the colors red (set \textit{B}) and yellow (set \textit{C}) for the positive chiral charge, while the blue (set \textit{B}) and purple (set \textit{C}) represent the negative Weyl chiralities. The sum of all chiral charges in the entire Brillouin zone vanishes, in agreement with the Nielsen-Ninomiya theorem~\cite{Nielsen1983}. One of the key properties of Weyl semimetals is the Berry curvature. Unlike the 2D case, which is described by a delta peak in the position of Weyl crossings,~\cite{Lopes2024} in a 3D system the Berry curvature is given by $ \boldsymbol{\Omega} = \pm{\boldsymbol{ k}}/{2k^{3}}$. This allows to visualize the Weyl points with positive or negative Weyl chirality as a source or drain of Berry curvature, respectively. To illustrate this, we show in Fig.~\ref{fig:WPS_berry}(a) the Berry curvature plot for a given plane of constant $k_z$. We also compute the evolution of Wannier charge centers (WCCs) for the selected crossing points, which further confirms the nontrivial behavior [Figs.~\ref{fig:WPS_berry}(c) and \ref{fig:WPS_berry}(e)].

\begin{figure*}
    \centering
    \includegraphics[width=\textwidth]{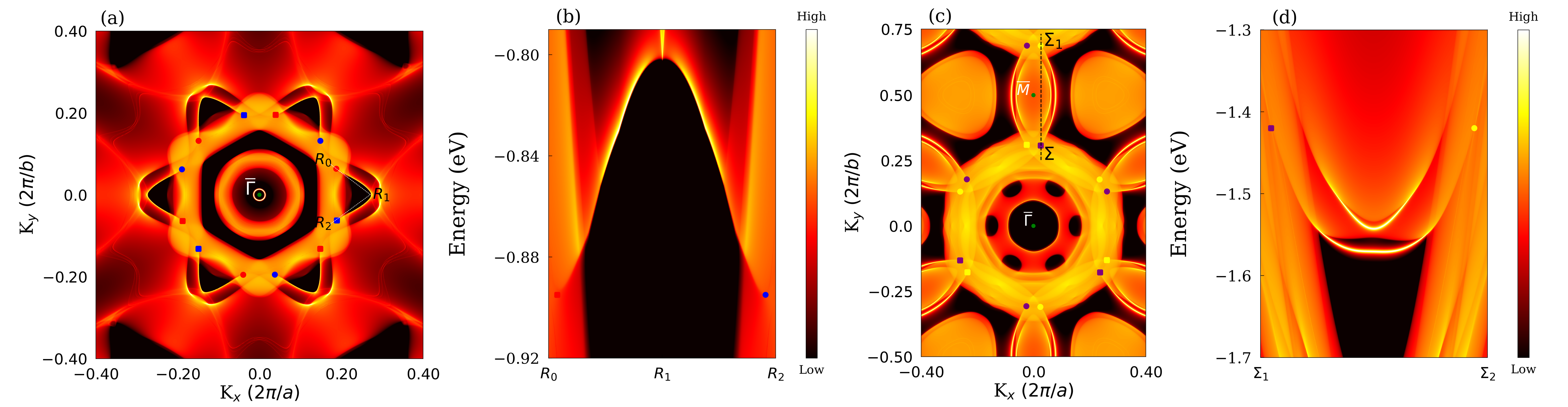}
    \caption{(a) Momentum-resolved electronic density of states for the energy cut at -0.89 eV on the (001) suface, for set $B$ of Weyl points. (b) Energy-resolved dispersion along the $R_0$-$R_1$-$R_2$, indicated by the white line in (a). In (c) and (d) are the momentum- and energy-resolved plots for the set $C$ of Weyl points, with a fixed cut at -1.41 eV.}
    \label{fermiarc}
\end{figure*}

To further confirm the topological properties of NiTe$_2$, we investigate its bulk-boundary correspondence by building a sufficiently large slab. Using a slab of 35 stacked unit layers and a distortion of $\Delta {z} = 5\%$, we search for the Fermi arcs. The topological surface states were investigated on the (001) surface in both ways, energy- and momentum-resolved density of states. Figures~\ref{fermiarc}(a) and \ref{fermiarc}(c) show the momentum-resolved electronic density of states for a fixed energy cut at $-0.89$ eV and $-1.41$ eV below the Fermi level, respectively. These values represent the energy cut at which the WPs from sets $B$ and $C$ emerge. The red and yellow (blue and purple) symbols in Figs. \ref{fermiarc}(a) and \ref{fermiarc}(c) represent the WPs with positive (negative) chirality of sets \textit{B} and \textit{C}, respectively. In both systems, one surface state emerges at the bulk-projected WP, forming a non-closed curve that connects to its respective opposite chiral pair.

The energy-resolved plots [Figs.~\ref{fermiarc}(b) and \ref{fermiarc}(d)] show the band dispersion along the $R_0$-$R_1$-$R_2$ and $\Sigma$-$\Sigma_1$ paths, as indicated in Figs.~\ref{fermiarc}(a) and \ref{fermiarc}(c), respectively. For set $B$ of WPs, one high intensity surface state can be seen at the interface with the bulk states. By means of energy-resolved density of states, this state can be seen emerging at the bulk WP projection on the surface, and showing a parabolic dispersion until it reaches its respective pair. For the set $C$ of Weyl crossings, the high intensity topological surface state shows an energy dispersion around 0.15 eV, achieving the minimal point in $-1.57$ eV, as shown in Fig.~\ref{fermiarc}(d).

In summary, based on first-principles and topological invariant calculations, we show a process for exploring Weyl semimetals by controlling the symmetry breaking in 1T-NiTe$_2$. We show that the predicted Dirac semimetal in NiTe$_2$ splits into a pair of type-II Weyl points by breaking the inversion symmetry, since the time-reversal symmetry is preserved. The inversion symmetry control has been performed by breaking the planar mirror $\sigma_{xy}$ symmetry, which could be experimentally performed by applying an electric field along the $z$ direction, or just by the effect of an appropriate substrate. Our results reveal additional sets of Weyl points, beyond those originating from the Dirac semimetal, coming out of gapped regions of the first Brillouin zone. The emergence of those unexpected sets of Weyl crossings is dependent on the weight of the symmetry breaking. For any symmetry breaking, the first set coming from the Dirac semimetal is always present near the Fermi level. Under a certain breaking weight, an additional set of Weyls emerges around -0.89~eV, and for a further strength, another set of Weyls is revealed around -1.41~eV from the Fermi level. Both additional sets present 6 pairs of Weyl nodes, the former being type-I while the second is type-II. This procedure for manipulating and creating Weyl semimetals, resulting in on/off of topological surface states, can be suitable for applications in Weyltronics.\\

See the supplementary material for the average potential energy curves of 1T-NiTe$_2$ under external effects [Fig. S1].\\

The authors acknowledge CAPES (Financial Code 001), CNPq, FAPEMIG and INCT-Nanocarbono for the financial support.  The authors also acknowledge LNCC (project SCAFMat2) and CENAPAD-SP (project proj483) for the computational time.

\section*{AUTHOR DECLARATIONS}
\subsection*{Conflict of Interest}
The authors have no conflicts to disclose.

\section*{DATA AVAILABILITY}

The data that support the findings of this study are available within the article and its supplementary material.

\bibliography{bibliography}

\end{document}